\documentclass[%
 reprint,
 amsmath,amssymb,
 aps,pre,twocolumn,
 superscriptaddress,
]{revtex4-2}

\usepackage{graphicx}
\usepackage{bm}
\usepackage{comment}
\usepackage{xcolor}
\usepackage{array} 
\usepackage{xr}

\newcolumntype{C}[1]{>{\centering\arraybackslash}p{#1}}

\newcommand{\bs}[1]{\boldsymbol{#1}}

\begin{document}

\title{Effect of swimming mode on shielding of odor traces in turbulence}

\author{Martin James}
\email{martin.james@edu.unige.it}
\affiliation{Machine Learning Genoa Center (MaLGa) \& Department of Civil, Chemical and Environmental Engineering, University of Genoa, Genoa, Italy}

\author{Francesco Viola}
\affiliation{Gran Sasso Science Institute (GSSI), L’Aquila 67100, Italy}
\affiliation{INFN–Laboratori Nazionali del Gran Sasso, Assergi, Italy}

\author{Agnese Seminara}
\email{agnese.seminara@unige.it}
\affiliation{Machine Learning Genoa Center (MaLGa) \& Department of Civil, Chemical and Environmental Engineering, University of Genoa, Genoa, Italy}
 
\begin{abstract} 

Marine organisms manipulate their surrounding flow through their swimming dynamics, which affects the transport of their own odor cues. We demonstrate by direct numerical simulations how a group of swimmers, moving at intermediate Reynolds numbers, immersed in a turbulent flow, alter the shape of the odor plume they release in the water. Odor mixing is enhanced by increased velocity fluctuations and a swimmer-induced flow circulation which widens the odor plume at close range while speeding up dilution of the chemical trace. Beyond a short-range increase in the likelihood of being detected, swimming considerably reduces detections with effects that can persist at distances of the order of ten times the size of the group or more. We find that puller-like swimmers are more effective at olfactory shielding than pusher-like swimmers. We trace this difference back to the dynamics at the swimmer location, which tends to trap odor at the source for pushers and to dilute it for pullers. Olfactory shielding is robust to changes in the conditions, and is more pronounced for weak turbulent Reynolds numbers and large swimmer Reynolds numbers. Our results suggest that olfactory shielding may play a role in the emergence of different swimming modalities by marine organisms. 

\end{abstract} 

\date{May 1, 2026}

\maketitle

\section{Introduction}

\begin{figure*}
\includegraphics[width=1\linewidth]{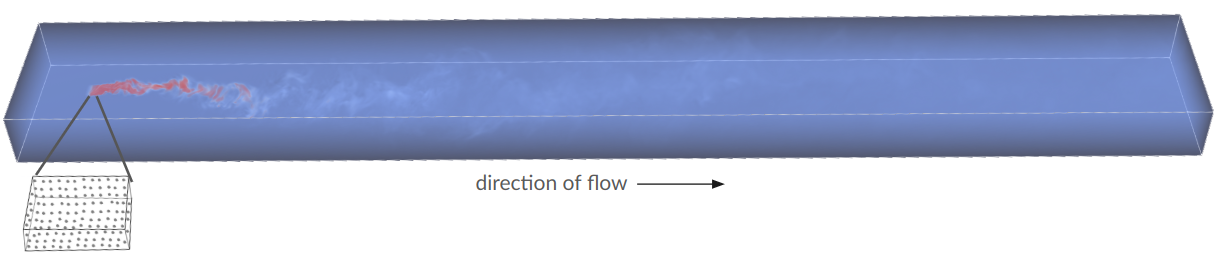} 
\caption{\label{fig:snapshot} A snapshot of the odor field emitted by the swimmers. The swimmers are placed in a cuboidal box of dimension $100\eta\times100\eta\times25\eta$ near the inlet, as shown in the inset (here $\eta$ is the Kolmogorov length scale). The simulation box has a dimension of $8000\eta\times1000\eta\times400\eta$.}
\end{figure*}

% OLFACTORY SHIELDING
Groups of swimming organisms are a major target for large predators who employ multiple sensory cues to track them down. In particular, swimmers give out their whereabouts by shedding odor, which is carried downstream of the group by water flow and can be detected by predators even from large distances (see, e.g.,~\cite{webster2009hydrodynamics,reddy2022olfactory} and references therein). As odor is transported away from the group, it undergoes turbulent mixing, which merges odor-laden and odor-free water. Mixing will ultimately erase their chemical trace as odor gets diluted below the threshold of detection of potential predators. The act of swimming in itself enhances turbulent mixing by actively manipulating the surrounding flow in a process known as biomixing, as documented by laboratory and numerical studies (see, e.g.,~\cite{kunze2019biologically} and references therein). Biomixing may even affect the global ocean circulation, although its quantification is challenging and hotly debated~\cite{katija2009viscosity,visser2007biomixing}. 
As a direct consequence of the documented enhanced mixing one expects that odor will dilute more efficiently, speeding up the elimination of the swimmers' own odor trace, as previously hypothesized for filter feeding by sessile organisms~\cite{delavan2012predator,alvarez2018modeling}. Given the fascinating diversity of species in the ocean, one naturally wonders whether specific ways of swimming may be more efficient than others at diluting their odor trace, which we refer to as olfactory shielding. Evidence that swimming kinematics can indeed shape chemical cue transport already exists; for example, studies have shown that undulatory swimming patterns can enhance the dispersal of odor cues~\cite{kamran2024does}.

Swimming organisms leverage a diversity of self-propulsion mechanisms, e.g.~beating or rotating appendages like cilia or flagella or propagating shape deformations~\cite{guasto2012fluid,wu2011fish}. 
Swimmers such as copepods, amphipods and other small crustaceans which move at intermediate Reynolds number ($Re$) form an important part of the marine ecosystem~\cite{kiorboe2009mechanistic}. In this manuscript we refer to them as ``mesoscale'' swimmers: note that this use of the term mesoscale is unrelated to its use in oceanography, where mescoscales are intermediate spatial scales. Mesoscale swimmers serve as pivotal food sources for larger marine species and play an important role in marine ecology~\cite{ratnarajah2023monitoring}. Such species are often found in large groups~\cite{flierl2015copepod}, inhabit regions of mild-to-moderate turbulence~\cite{michalec2015turbulence,omori1982patchy,guasto2012fluid} and show a fascinating variety in behavior~\cite{ardeshiri2016lagrangian,ardeshiri2017copepods,mousavi2024efficient,qiu2022navigation}. Field and laboratory observations suggest that several species of crustaceans form dense swarms composed of thousands of individuals that can maintain their position in mild turbulence~\cite{buskey1996swarming,buskey1995role,emery1968preliminary,ueda1983underwater}. For instance, \textit{Dioithona oculata} swarms withstand currents up to 2 cm/s (fluid $Re > 4000$)~\cite{buskey1996swarming,buskey1995role,ambler2002zooplankton}, while several mysid species aggregate in shallow surge channels during day and night~\cite{emery1968preliminary,ohtsuka1995direct}.

Swimmers moving at low and intermediate Reynolds number can exhibit pusher or puller dynamics, depending on whether they generate thrust at their rear or front during specific swimming modes~\cite{tack2024ups,lauga2009hydrodynamics}. The hydrodynamic interactions by pushers and pullers generate distinct patterns in their surrounding water flow, both individually~\cite{kiorboe2014flow,van2003escape,lauga2009hydrodynamics} as well as in groups~\cite{zantop2022emergent,cavaiola2022swarm}. But whether and how mesoscale pushers \emph{vs} pullers are more or less efficient at olfactory shielding remains to be understood. 

In this article, we provide a detailed analysis of the coupling of mechanical and chemical signals from mesoscale swimmers in shallow, turbulent flows, drawing inspiration from the collective dynamics of crustacean swarms~\cite{buskey1996swarming,buskey1995role,emery1968preliminary}. We conduct high-resolution direct numerical simulations using a model system of a collection of point swimmers in an open channel turbulent flow. %we explore the propagation and interaction of these signals. 
We show that groups of swimmers enhance mixing, and puller-like swimmers are more efficient than pusher-like swimmers at olfactory shielding. Furthermore, we examine how swimmer kinematics and positional noise influence olfactory shielding. We find that puller-like swimmers remain more effective even under high positional noise. We also demonstrate that olfactory shielding is stronger for large swimmer Reynolds numbers and low flow Reynolds numbers, suggesting that olfactory shielding may contribute selective pressure on mesoscale swimmers in the ocean.

\section{Modeling and methodology}
\label{sec:methods}

We consider a collection of mesoscale swimmers, embedded in an open channel flow. The fluid flow is modeled through direct numerical simulation of the incompressible Navier-Stokes equation

\begin{table}[h]
\centering
\begin{tabular}{|C{2cm}|C{2cm}|C{2cm}|}
\hline
No. & $Re_\tau$ & $Re_s$ \\
\hline
1 & 560 & 50 \\
2 & 560 & 25 \\
3 & 560 & 10 \\
4 & 685 & 50 \\
5 & 395 & 50 \\
6 & 225 & 50 \\
\hline
\end{tabular}
\caption{\label{table:simulation_list} List of environments with the corresponding friction Reynolds numbers $Re_\tau$ of the channel flow and the swimmer Reynolds numbers $Re_s$.}
\end{table}

\begin{align}
\label{eq:ns}
    \rho\left(\frac{\partial \bs{u}}{\partial t} + \bs{u} \cdot \nabla \bs{u}\right) &= -\nabla P + \mu \nabla^2 \bs{u} + \bs{f} + \bs{f}_p; \\
    \nabla\cdot\bs{u}&= 0. \nonumber
\end{align}
Here, $\bs{u}$ is the fluid velocity field, $P$ is the pressure, $\rho$ is the fluid density, $\mu$ is the dynamic viscosity and $\bs{f}_p$ is the forcing due to the swimmer dynamics. We solve Eq.~\eqref{eq:ns} using a second-order central finite-difference scheme on a staggered grid~\cite{viola2020fluid}.

\begin{figure*}
\includegraphics[width=0.8\linewidth]{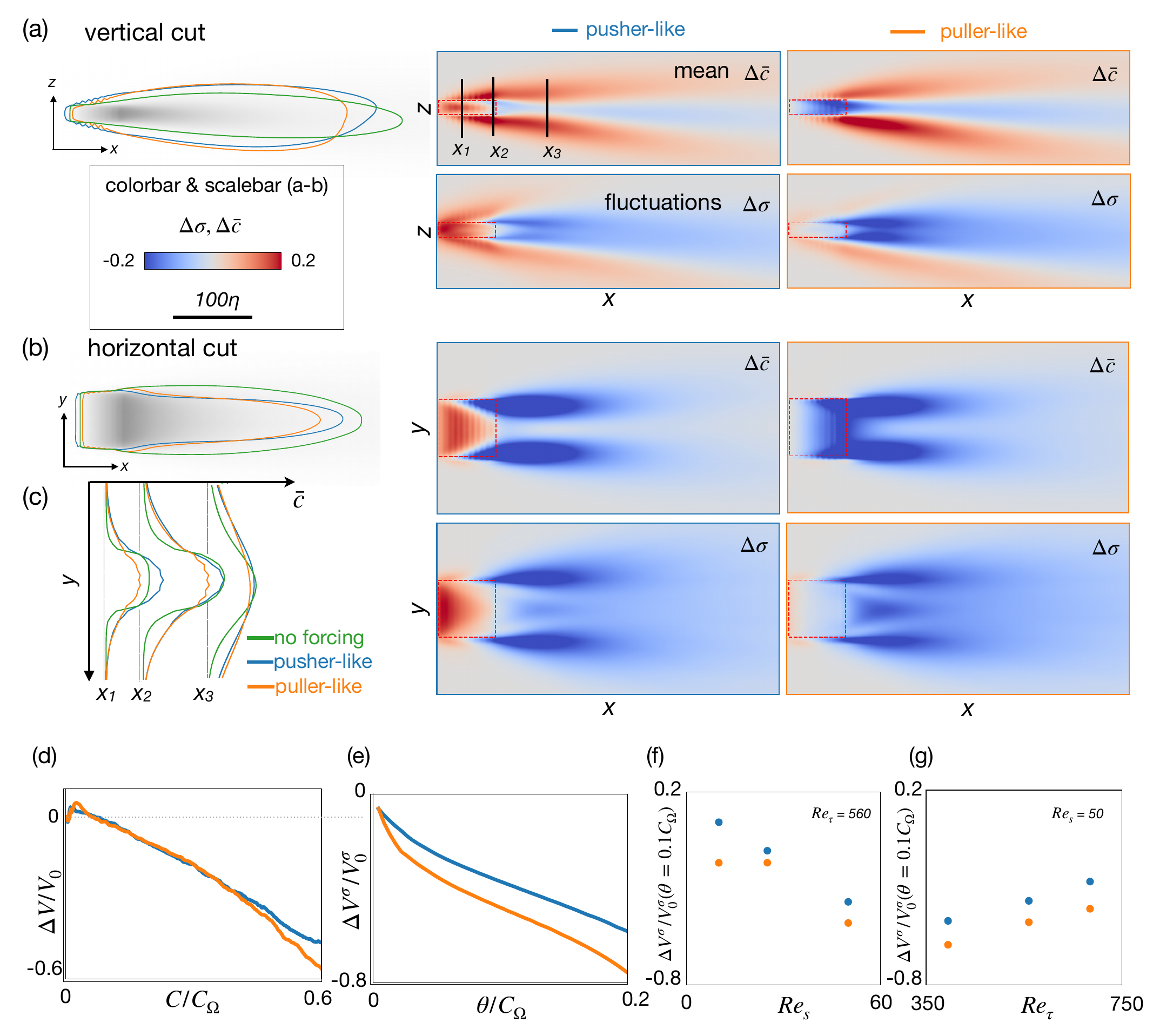} 
\caption{\label{fig:mean} Statistics of the mean odor field. The contour plots show how the range of the mean odor field varies between pusher-like, puller-like and neutral swimmers for vertical and horizontal mid-plane cross-sections. (a) Vertical cross-sections of the mean odor differential $\Delta \bar{c}(\bs{x})$ for pusher-like swimmers and puller-like swimmers and the corresponding odor standard deviation differential $\Delta \sigma(\bs x)$. The region containing the swimmers is marked by a red box. (b) The corresponding figures for the horizontal cross-section. (c) Odor profiles at different distances from the swimmers' location. (d) Shielding intensities of the mean odor field $\Delta V(C)/V_0(C)$. (e) Shielding intensity of the odor fluctuations $\Delta V^\sigma(\theta)/V^\sigma_0(\theta)$ for pusher-like and puller-like swimmers, showing that the swimming dynamics dampens large fluctuations. (f) Impact of swimmer Reynolds number and (g) flow Reynolds number on the shielding intensity of the odor fluctuations $\Delta V^\sigma(\theta)/V^\sigma_0(\theta)$ for pusher-like and puller-like swimmers (evaluated at $\theta = 0.1C_\Omega$).} 
\end{figure*}

The channel is forced with a constant pressure gradient $\bs{f}$~\cite{quadrio2016does}. The boundary conditions for the channel flow are as follows: a fixed velocity on the bottom boundary (to represent the appropriate swimming velocity of the swimmers, as the simulations are carried out in the reference frame of the swimmers), free slip on the top boundary ($u_z=0, \partial_zu_x=\partial_zu_y=0$) and periodic in the other directions. We simulate three different fluid friction Reynolds numbers $Re_\tau = u_\tau H/\nu$, where $u_\tau$ is the shear velocity, $H$ is the channel height and $\nu$ is the kinematic viscosity ($\nu = \mu/\rho$), as shown in Table~\ref{table:simulation_list}. We choose Environment 1 in Table~\ref{table:simulation_list} as our base environment and all units are normalized by using the Kolmogorov scales in this simulation. The simulation box has a dimension of $8000\eta\times1000\eta\times400\eta$, where $\eta$ is the Kolmogorov length scale.
 
The odor field is modeled using an advection-diffusion equation.

\begin{equation}
\label{eq:odor}
    \frac{\partial c}{\partial t} + \bs{u} \cdot \nabla c= D \nabla^2 c + q.
\end{equation}
Here $c$ is odor concentration, $D$ is the diffusivity and $q$ is the odor source. We assume perfect adsorption on the bottom boundary ($c=0$), no flux at the top boundary ($\partial_zc=0$) and outflow conditions at the other boundaries. The source term $q$ at the position of the swimmers model a constant odor leakage rate. For our simulations, we use Schmidt number $Sc=1$. Although $Sc$ in water is typically higher than this, we anticipate a weak dependence on $Sc$ since we are interested in large-scale statistics~\cite{rigolli2022learning,selander2020chemical}.

\begin{figure*}
\includegraphics[width=0.8\linewidth]{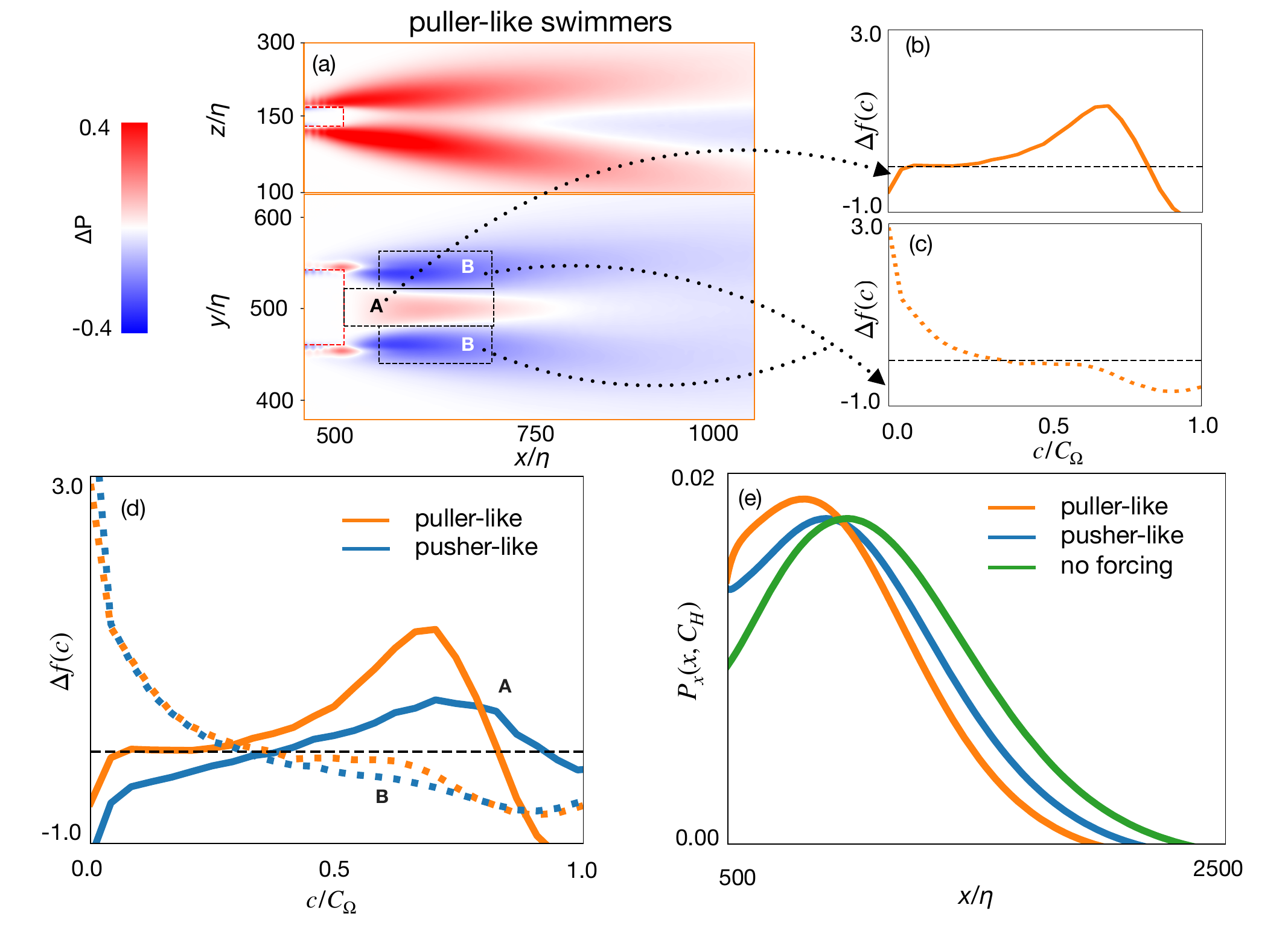} 
\caption{\label{fig:probability} The effect of swimming dynamics on the detection probability. (a) $\Delta P(\bs x; C_H)$ for puller-like swimmers. (b) and (c) Odor pdf differential $\Delta f(c)$ in regions A and B, respectively, for puller-like swimmers. (d) Odor pdf differential $\Delta f(c)$ for both types of swimmers; statistics in regions A and B are represented with solid and dotted lines respectively. (e) Detection probability $P_x(x;C_H)$ as a function of stream-wise distance, averaged along $y$ and $z$. The $x-$axis starts at $x=500$, where the group of swimmers ends. At long distances $x\gtrsim 700\,\eta$ the detection probability is reduced due to the swimming dynamics. Shielding persists up to $x\sim 2500\,\eta$.}
\end{figure*}

To model swimmer kinematics, we consider two different settings. In the first, swimmers maintain their relative positions within the group throughout the simulation, effectively forming a fixed grid in the reference frame. In the second setting, swimmer kinematics are modeled using a stochastic process, as explained later. 
To model the forcing $\bs{f}_p$ due to the swimmers, note that the flow field due to neutrally buoyant micro and mesoscale swimmers can be approximated by a force dipole~\cite{jiang2004hydrodynamics,drescher2011fluid,wagner2014mixing}. Further addition of torque dipoles combined with rigid boundary conditions can effectively capture swimmers moving at even higher $Re$~\cite{ventejou2024universal}. However, since our swimmers are of the order of the Kolmogorov scale and move at relatively low $Re$, we model our swimmers using a force dipole approximation. Each swimmer is modeled as a force dipole $F(\delta(\bs{x}+\bs{r})-\delta(\bs{x}-\bs{r}))\hat{\bs{x}}$, where $r = 2.5\eta$ is half the length of the swimmer and $F$ is the magnitude of the forcing. The dipole axis is aligned with the mean relative flow field, emulating the rheotactic orientation observed in crustacean swarms within shallow flows~\cite{buskey1995role,buskey1996swarming,emery1968preliminary}. The orientation of the force dipole determines whether the swimmer is pusher-like or puller-like. When the forcing points outwards relative to the swimmer's axis, the swimmer ``pushes'' the fluid away and is a pusher-like swimmer whereas, when it points inwards, it ``pulls'' the fluid in, thus modeling a puller-like swimmer~\cite{lauga2009hydrodynamics}. The magnitude of $F$ is computed based on the experimental results for the drag force at intermediate Reynolds numbers~\cite{granata1991fluid} (also see Eq.~\eqref{eq:drag} in the Supporting Information~\cite{supplemental}). For the first set of simulations, the swimmers are placed in a fixed grid relative to the frame of reference in a cuboidal box of dimension $100\eta\times100\eta\times25\eta$ close to the inlet with a total of $10 \times 20 \times 5$ swimmers along the streamwise, spanwise and wall-normal directions respectively. 

To evaluate the effect of the kinematics of the swimmers on odor shielding, we conduct additional simulations where the swimmers are not fixed on a grid, but move according to the following Ornstein-Uhlenbeck process:
\begin{equation}
    d\bs{x}_i = -\frac{1}{\tau}(\bs{x}_i-\bs{\mu}_i)dt + \sigma_{OU} \sqrt{dt}\bs{\eta_i},
\end{equation}
where $\bs{x}_i$ is the position of $i$th swimmer, $\bs{\mu}_i$ its equilibrium position, $\tau$ is the relaxation time, $\bs{\eta_i}$ the noise and $\sigma_{OU}$ is the strength of the noise. All three components of $\bs{\eta_i}$ are drawn from Gaussian distributions with zero mean and unit standard deviation. The equilibrium position of the swimmers are assumed to be the same as in the fixed grid configuration described above. However, unlike in the previous configuration, the swimmers can disperse, with $\sigma_{ms} = \sigma_{OU}\sqrt{\tau/2}$ defining the standard deviation of the swimmers' position relative to their equilibrium positions.

We consider three different Reynolds numbers for the swimmers $Re_s = u_sD/\nu$, where $u_s$ is the magnitude of the swimming velocity and $D$ is the diameter, as shown in Table~\ref{table:simulation_list}. To achieve different swimmer Reynolds numbers, the velocities of the swimmers are adjusted by modifying the fluid velocity boundary condition applied on the bottom wall, with all other parameters held constant. The simulations are conducted in the reference frame of the swimmers. All simulations are run until at least $T=4 \times 10^5 \tau_\eta$ to reach a statistically steady state before the results are evaluated.

The list of environments is given in Table~\ref{table:simulation_list}. All figures correspond to the base environment (Simulation 1 in Table \ref{table:simulation_list}) unless otherwise mentioned. For each of these environments, we conduct three distinct simulations: In the first two, we incorporate the dynamics of the swimmers by accounting for the forcing imposed by the swimmers on the fluid -- pusher-like or puller-like; then we remove this forcing, providing a baseline against which to measure the swimmers' effects. In the baseline simulation, the swimmers are retained as passive `ghosts' that do not influence the flow, but emit odor at the same rate as the other two cases. The simulation setup is shown in Fig.~\ref{fig:snapshot}, illustrating a typical case of odor dispersion in a channel flow. 

\begin{figure*}
\includegraphics[width=1\linewidth]{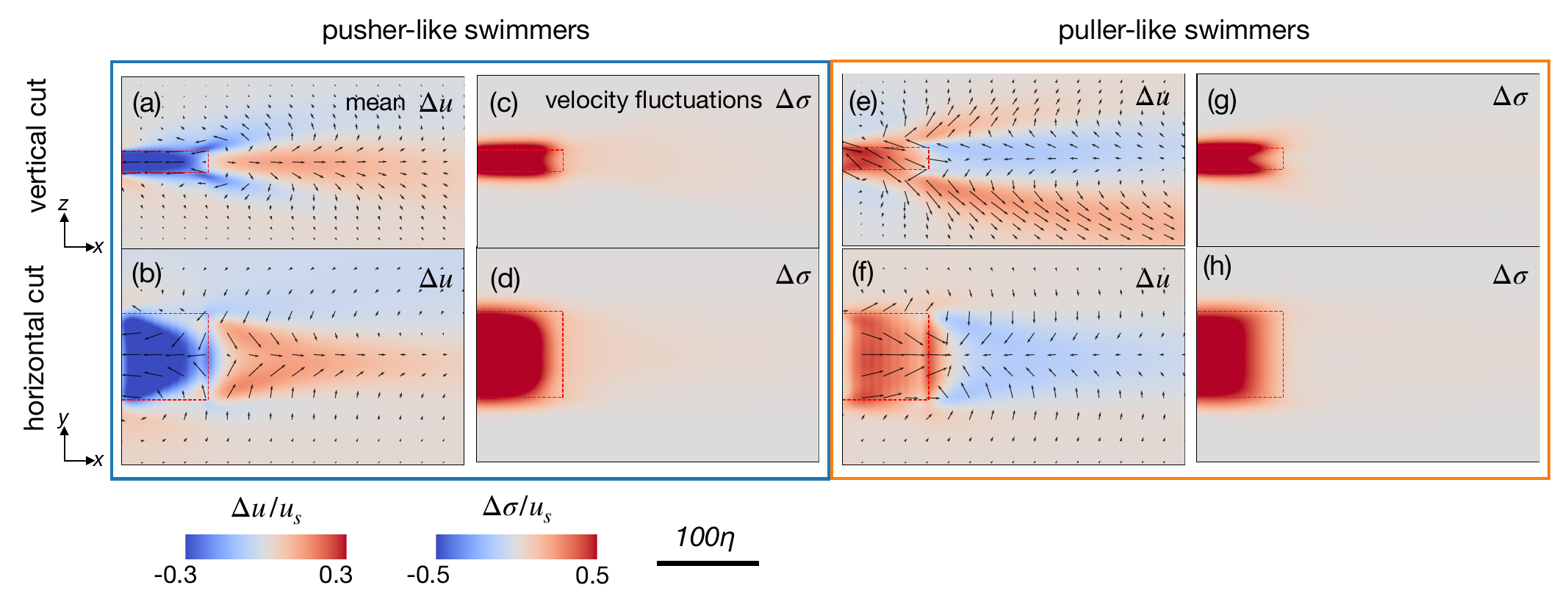} 
\caption{\label{fig:velocity} Effect of swimming dynamics on fluid velocity. Difference in the mean streamwise velocity due to the swimmers' dynamics, normalized with the swimmer speed for (a,b) pusher-like and (e,f) puller-like swimmers. Difference in the standard deviation of the cross-stream velocity fluctuations for (c,d) pusher-like and (g,h) puller-like swimmers.}
\end{figure*} 

\section{Results}
To determine whether swimmers' dynamics leave their signatures in the odor distribution over long time-scales, we begin our analysis by examining the effect of hydrodynamic interactions on the mean and standard deviation of the odor field (Fig.~\ref{fig:mean}). In this first part, we limit our analysis to swimmers fixed on a grid relative to the frame of reference, disregarding the effect of their kinematics. 
Since the odor distribution is inhomogeneous and statistically stationary, we approximate the ensemble average of the mean and standard deviation of the odor field with temporal averages:
\begin{align}
\label{eq:odor_mean}
    \bar{c}(\bs x) &= \langle c(\bs x,t)\rangle \\
    \sigma(\bs x) &= \langle (c(\bs x,t) - \bar{c}(\bs x,t))^2\rangle^{\frac{1}{2}}. \nonumber
\end{align}
To account for the effect of the swimmers, we now define differential quantities by comparison with the baseline simulation that ignores the presence of the swimmers:

\begin{align}
\label{eq:odor_mean_diff}
    \Delta \bar{c}_s(\bs x) &= [\bar{c}_s(\bs x) - \bar{c}_0(\bs x)]/C_\Omega\ \\
    \Delta \sigma_s(\bs x) &=[\sigma_s(\bs x) - \sigma_0(\bs x)]/C_\Omega, \nonumber
\end{align}
where the subscript $s$ stands for swimmers (either pusher-like or puller-like), and the subscript 0 represents the baseline simulation. Mean and fluctuations are normalized by $C_\Omega = \text{max}(\bar{c}_0)$, which occurs within the volume $\Omega$ containing the swimmers.

Swimmer dynamics strongly affect the odor mean and fluctuations, with deviations up to 20\% of $C_\Omega$ (Fig.~\ref{fig:mean}~(a-b)). This effect varies in different regions of space, as swimmers are clustered together. Within the region $\Omega$ that contains the swimmers, pusher-like and puller-like dynamics affect odor in opposite ways. Pusher-like swimmers  
\emph{concentrate} the odor average within $\Omega$ and \emph{increase} odor fluctuations, whereas puller-like swimmers \emph{dilute} the odor average and \emph{decrease} its fluctuations (Fig.~\ref{fig:mean}~(a-b)). 
Outside of $\Omega$, both swimming modalities have the same qualitative effects, which however differ quantitatively and vary in space. Within the wake, both swimmers lead to a decrease in odor mean and fluctuations; outside of the wake, in the wall-normal direction -- i.e.~vertical cut, Fig.~\ref{fig:mean}~(a), above and below the wake -- swimmers increase odor mean and fluctuations; whereas in the spanwise direction --i.e.~horizontal cut Fig.~\ref{fig:mean}~(b), swimmers decrease odor mean and fluctuations.

\begin{figure*}
\includegraphics[width=0.8\linewidth]{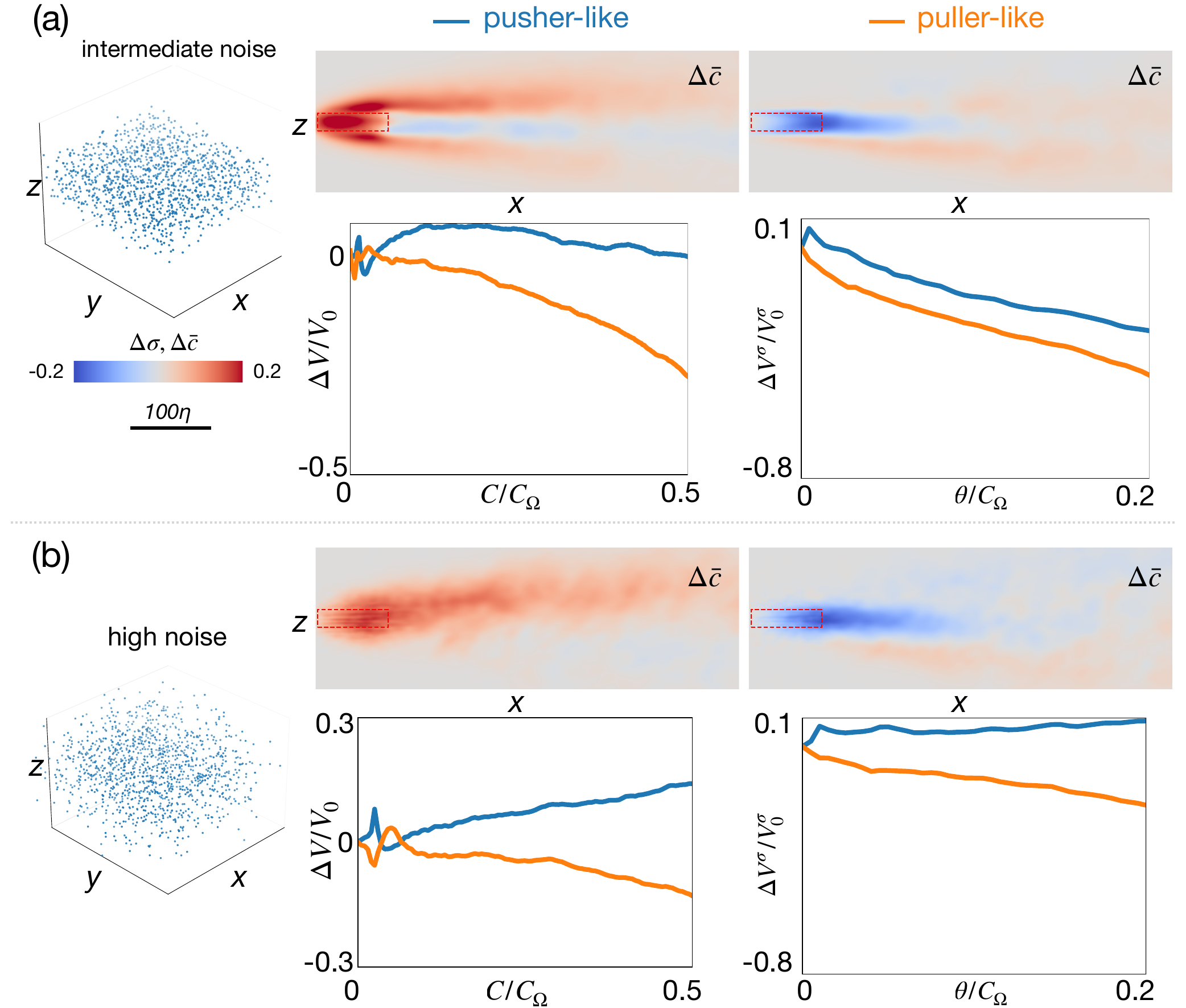} 
\caption{\label{fig:kinematics_both}  Effect of swimmer kinematics on odor shielding under (a) intermediate positional noise (top row) and (b) high positional noise (bottom row), with snapshots of the swimmer positions is shown on the left.} 
\end{figure*}

To quantify the overall effect of swimmers in the entire volume, we define the volumes $V(C)$ where the mean odor concentration exceeds $C$ and $V^\sigma(\theta)$ where the fluctuations exceed $\theta$:
\begin{align}
\label{eq:odor_vol}
    V_s(C) = \int H(\bar{c}_s(\bs x) - C)\text{d}{\bs x}, \nonumber \\
    V_s^\sigma (\theta) = \int H(\sigma_s(\bs x) - \theta)\text{d}{\bs x}, \nonumber 
    \end{align}
where $H(x)$ is the Heaviside step function and %$\Omega$ is 
integration is extended to the entire domain. We define the shielding efficiency as the decrease in the volumes defined above:
    \begin{align}
    \Delta V_s (C) = V_s (C) - V_0 (C), \nonumber \\
    \Delta V_s^\sigma (\theta) = V^\sigma_s (\theta) - V^\sigma_0 (\theta). \nonumber 
\end{align}
When $C\approx 0$, swimming dynamics has little to no effect (or even a weak negative effect) in shielding the mean odor, with puller-like swimmers performing slightly worse than pusher-like ones (Fig.~\ref{fig:mean}~(d)). As $C$ is increased, $\Delta V (C) < 0$ thus both types of swimmers decrease the total volume where the average concentration is above the threshold. For a threshold of $C/C_\Omega=0.5$, pullers shrinks $V(C)$ by $50\%$, while pushers shrink $V(C)$ by $45\%$. Similarly, shielding efficiency for odor fluctuations (Fig.~\ref{fig:mean}~(e)) shows that regions with extreme fluctuations are significantly reduced when the swimmer dynamics are taken into account. Interestingly, puller-like swimmers exhibit better shielding of both mean and fluctuations, which may reduce the chances of detection. These results stand in contrast to that of a single swimmer, which yields no significant shielding effect (Fig.~\ref{fig:single_si}).

Having examined odor shielding in the base environment, we now evaluate how the shielding changes as the Reynolds numbers of the swimmers and the advecting fluid are changed. At higher swimmer Reynolds numbers, the effect of forcing by the swimmers is expected to be more significant leading to an increase in shielding efficiency with $Re_s$. Our numerical experiments confirm this expectation (Fig.~\ref{fig:mean} (f) and Fig.~\ref{fig:reynolds}). At lower swimmer $Re_s$ of 10 and 25, the swimmers' dynamics have little effect on shielding the mean odor (Fig.~\ref{fig:reynolds} (a)). However, the fluctuations in the odor field do carry signatures of the swimming dynamics even at lower $Re$ for puller-like swimmers (Fig.~\ref{fig:mean} (f) and Fig.~\ref{fig:reynolds} (b)). To understand how the fluid $Re$ affects shielding, we repeat our analysis for two more fluid $Re$. Our analysis indicates that the effect of odor shielding is more prominent at lower fluid Reynolds numbers (Fig.~\ref{fig:mean} (g)). What is critical, however, is that puller-like swimmers continue to outperform pusher-like swimmers in all three fluid $Re$ considered in our analysis. A comparison of the swimmer and fluid Reynolds numbers that we consider with field observations and experiments of copepod swarms reported in Ref.~\cite{buskey1996swarming} is provided in Table~\ref{table:reynolds_comparison}.

We hypothesize that the damping of odor mean and fluctuations is a consequence of enhanced mixing due to the presence of the swimmers. To corroborate this intuition, we examine the full probability distribution of the odor field: 
\begin{equation}
   f(c, \bs x) = \langle \delta(c(\bs x, t) - c) \rangle, 
\end{equation}
where $f(c, \bs{x})$ represents the probability density of the odor field $c(\bs{x}, t)$ taking the value $c$ at a specific point $\bs{x}$. Here, $\delta(\cdot)$ is the Dirac delta function and the angular bracket denotes the expectation. The right tail of $f(c, \bs x) $ defines the probability of detecting odor at a threshold $C$ through:
\begin{equation}
    P(\bs x, C) = \int_C^\infty f(c, \bs x)\text{d}c.
\end{equation} 
 To quantify shielding we focus on the variation of $P$ due to the swimmers relative to the baseline simulation: $\Delta P_s(\bs x, C) = P_s(\bs x, C) - P_0(\bs x, C)$. $C$ is often hard to measure as it depends on a complex interaction between environmental conditions, the chemical identity of the odor and the predator. We exemplify results using two thresholds, ``high'' $C_H = 0.1\,C_\Omega$ and ``low'' $C_L = 0.01 \,C_\Omega$.

The probability of detection generally increases above and below the swimmers, and it decreases left and right (vertical and horizontal cross cut, Fig.~\ref{fig:probability}~(a-b)).
Within the horizontal cut, a further distinction needs to be made between a thin core region in the immediate downstream of the group (region A in Fig.~\ref{fig:probability}~(a)), where swimmers \emph{increase} their detection probability, and regions outside this thin core, both further downstream and at the outer edges of the wake (region B in the Figure), where swimmers \emph{decrease} their detection probability. 
The same qualitative pattern holds for lower thresholds, with region A expanding further and further downstream as the threshold is decreased (see Fig.~\ref{fig:prob_full}), consistent with the results illustrated for the mean (Fig.~2~(d-e)).
For pusher-like swimmers, all patterns are qualitatively preserved, although the intensity of shielding differs (see Fig.~\ref{fig:prob_full}).

\begin{figure*}
\includegraphics[width=0.8\linewidth]{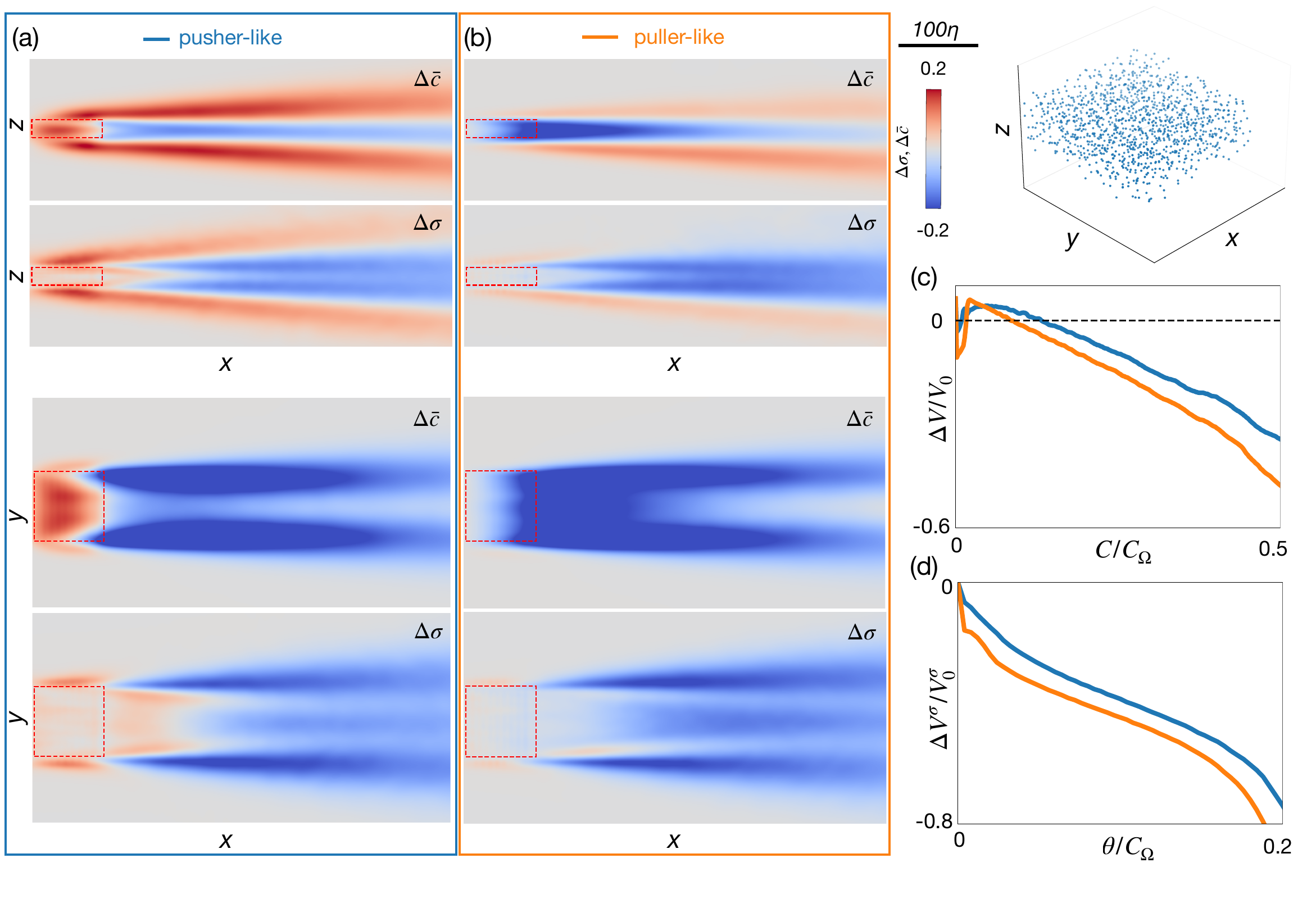} 
\caption{\label{fig:kinematics} Effect of swimmers' kinematics on odor shielding when their mean position remains stationary relative to the ground. A snapshot of the swimmers' positions is shown on the top right. Vertical and horizontal cross-sections of $\Delta \bar{c}(\bs{x})$ and $\Delta \sigma(\bs x)$ for (a) pusher-like swimmers and (b) puller-like swimmers. (c) Shielding intensities of the mean odor field $\Delta V(C)/V_0(C)$. (d) Shielding intensity of the odor fluctuations $\Delta V^\sigma(\theta)/V^\sigma_0(\theta)$ for pusher-like and puller-like swimmers. }
\end{figure*}

Unintuitively, the two swimmer types have opposite effects at the odor source, yet they both shrink their own detection probability at large enough distances from the group. This is best summarized by monitoring the probability of detection as a function of distance $x$ downstream of the swimmers: 
\begin{align}
    P_x(x, C) &= \frac{1}{L_zL_y}\int_{0}^{L_z}\int_0^{L_y}  P(\bs x, C)\text{d}y\text{d}z.
    %\\
    %\Delta P_x(x, C) &= P_{x_0}(x, C) - P_{x_s}(x, C)
\end{align}
The group increases their probability of detection at close range, i.e.~within  $\sim$~$200\eta$ from the end of the group, about twice the length of the group. At longer range the group decrease their detection probability (Fig.~\ref{fig:probability} (e)). Importantly, shielding  persists up to  $\sim$~$2000\eta$, i.e.~20 times the length of the group. To reconcile this unintuitive mismatch, we analyze the full probability density function within regions A and B  and compare the simulations with and without swimmers $\Delta f_s(c, \bs x) = f_s(c, \bs x) - f_0(c, \bs x)$. In region A the swimmer dynamics dampen both extremes of the odor distribution (Fig.~\ref{fig:probability}~(b)). This is a signature of increased mixing, which homogenizes the odor field towards its mean concentration. In region B the swimmer dynamics dilutes the odor, presumably by spreading it in a larger region further outward (Fig.~\ref{fig:probability}~(c)). Pusher-like swimmers behave qualitatively similarly (Fig.~\ref{fig:probability}~(d)). These signatures of enhanced mixing suggest that the velocity fluctuations due to the swimmer dynamics play a major role in odor transport. The effect of swimmer and flow Reynolds numbers on the probability of detection is shown in Fig.~\ref{fig:reynolds}. As we already observed for the mean odor field, shielding of the probability of detection is more effective at higher swimmer Reynolds numbers and lower flow Reynolds numbers.

To understand the origin of shielding and corroborate the role of increased mixing, we examine how swimmers affect their surrounding flow field (Fig.~\ref{fig:velocity}). First, there is a marked difference in how swimmers affect the mean flow at their own location. Pusher-like swimmers slow down the mean flow, whereas puller-like swimmers accelerate the mean flow (Fig.~\ref{fig:velocity}~(a-b)), consistent with the results on the motility of inertial pusher-like and puller-like squirmers~\cite{wang2012inertial}.
By odor conservation, a slow down of the mean flow by the pushers will tend to concentrate the odor within the source region $\Omega$, whereas the acceleration of the mean flow by the pullers will tend to dilute it. 

Downstream of the source, the mean flow pushes the water up and down vertically (Fig.~\ref{fig:velocity}~(a,b)) and draws it inward laterally in the horizontal plane (Fig.~\ref{fig:velocity}~(e,f)). As a result, odor spreads vertically and shrinks laterally, and the probability of detection increases outside of the wake in the vertical plane, whereas it decreases outside of the core region in the horizontal plane. 
Finally, increased mixing in the core region A (Fig.~\ref{fig:probability}~(d)) is caused by increased fluctuations, which occur for both kinds of swimmers (Fig.~\ref{fig:velocity}~(c-d,g-h)).

\begin{table}[h]
\centering
\begin{tabular}{|C{2.5cm}|C{3cm}|C{3cm}|}
\hline
Study & Swimmer Reynolds number $Re_s$ & Bulk Reynolds number of the flow $Re_b$ \\
\hline
Present work & 10 -- 50 & 6,000 -- 18,050 \\
\hline
Laboratory experiment in Ref.~\cite{buskey1996swarming} & $\leq$20 & $\leq$2,000 \\
\hline
Field observations in Ref.~\cite{buskey1996swarming} & $\leq$20 & $\leq$15,000 \\
\hline
\end{tabular}
\caption{\label{table:reynolds_comparison}Comparison of the swimmer Reynolds number $Re_s$ and the bulk Reynolds number $Re_b = U_b H/\nu$, where $U_b$ is the mean flow velocity and $H$ is the flow depth, between the present study and the experimental and field observations of copepod swarms reported in Ref.~\cite{buskey1996swarming}.}
\end{table}

In summary: both swimmer types cause the odor plume to bulge vertically and shrink laterally, and also increase velocity fluctuations. As a consequence, swimmers speed up dilution of the odor downstream of their position thus shrinking the probability that a potential predator may detect them. However, pullers are more effective at olfactory shielding because they additionally \emph{dilute} odor at the source, whereas pushers \emph{trap} odor at the source. Note that our problem is anisotropic due to the presence of the wall and the arrangement of the swimmers within the group, as noticed by comparing odor patterns in the spanwise and wall-normal directions (Fig.~2~(a-b)).
The anisotropy of the cuboid containing the swimmers is the dominant anisotropy. Indeed, when considering a cuboid with the same shape but rotated of $90$ degrees, we obtain a similar pattern as the one visualized in Fig.~2, with the $y-$ and $z-$ axis swapped (see Supplementary Fig.~\ref{fig:mean_vertical}). Small differences with respect to Fig.~2 can be attributed to the anisotropy of wall turbulence (see Supplementary Fig.~\ref{fig:vel_si}).
While the shape of the group of swimmers alters the quantitative aspects of the problem, odor shielding persists for the multiple swimmer aspect ratios that we present in the supporting information~\cite{supplemental} (see Fig.~\ref{fig:mean_same} and Fig.~\ref{fig:mean_vertical}). 

Having examined in detail the mechanism behind olfactory shielding, let us now evaluate the effect of the kinematics of the swimmers. Mesoscale swimmers such as copepods show significant positional noise within the swarm, as has been reported from field observations and experiments~\cite{buskey1995role,buskey1996swarming}. To account for this, we model noisy kinematics of the swimmers using an Ornstein-Uhlenbeck process as explained in Section~\ref{sec:methods}. We consider two different values for the positional noise, a first level of noise intensity (``intermediate'' $\sigma_{ms}\approx d$, where $d$ is the mean separation between the swimmers) and a vigorous noise intensity (``high'' $\sigma_{ms} \approx 5d$), with the swimmers embedded in Environment 1 (Fig.~\ref{fig:kinematics_both}). Comparing with Fig.~\ref{fig:mean}, we see that qualitatively the mean odor differential $\Delta \bar{c}(\bs{x})$ remains similar. 
Puller-like swimmers persistently shield the odor, with an efficiency that decreases as the positional noise is increased. Importantly, noise does not affect the difference between swimming modes, with pusher-like swimmers robustly less efficient than puller-like swimmers, so that at high noise intensity their shielding effect vanishes (Fig.~\ref{fig:kinematics_both} (b) and \ref{fig:kinematics_high_noise}).
This suggests that the ecological advantages for puller-like swimmers are robust and preserved regardless of the intensity of positional noise. \\
To further corroborate the results in an experimentally observed condition, we leverage the observations reported in Refs.~\cite{buskey1995role,buskey1996swarming,ueda1983underwater}, where swarms of copepods were observed to maintain their position relative to the ground both in the laboratory as well as in the field.  
To match more closely these observations, we conduct an additional set of simulations where the swimmers' mean position remains stationary relative to the ground, even within an incoming current of a few cm/s~\cite{ambler2002zooplankton}. To this end, we change the flow Reynolds number such that the average velocity of the swimmers, relative to the surrounding flow, remains the same as in Fig.~\ref{fig:kinematics_both}, but the the mean position of the swarm is fixed relative to the ground. In these simulations, the swimmers are embedded in Environment~6 and move with intermediate positional noise  ($\sigma_{ms} \approx d$). Our results show that, under these conditions, both $\Delta \bar{c}(\bs{x})$ and $\Delta \sigma(\bs x)$ are strongly shielded, with puller-like swimmers outperforming pusher-like swimmers (Fig.~\ref{fig:kinematics}). This further illustrates that olfactory shielding by mesoscale swimmers remains significant in experimentally relevant conditions, as long as the swarm moves relative to the incoming current and whether its absolute position relative to the seafloor is fixed or not.

In this work, we demonstrate how the hydrodynamic fluctuations introduced by a collection of mesoscale swimmers interact with their own odor field in a turbulent flow. We focus on two prototype swimmer types, pusher-like and puller-like, and show that in both cases the hydrodynamic interactions effectively erase the group's odor trace downstream of the swimmers. Importantly, puller-type swimmers shield odor more efficiently than pusher-like swimmers. 
Shielding is observed both in an idealized model where the relative positions of the swimmers are fixed on a grid, and in more a realistic kinematic model where the position of the swimmers is affected by noise. While noise intensity modulates the shielding efficiency, it does not affect the robust difference between swimming modalities. How the detailed statistics of the positional noise is generated by a dynamical model of swimming, and how it affects   shielding efficiency is left for future work.
Moreover, the effect increases at relatively low turbulence and fast swimming, providing potentially relevant mechanisms for predator avoidance and for larger and more vigorous swimmers. Further work is needed to test the ability of higher organisms to shield their own odor trace, and how it depends on the diversity of fascinating swimming modes observed in nature. Our results may be further extended to ask how swimming mode affects pheromone transport at the individual level, and in conditions that mimic pelagic lifestyle far from the seabed.

Interestingly, filter-feeding organisms like clams have been experimentally shown to frantically increase their motion upon exposure to predator cues~\cite{delavan2012predator}. These observations have been linked to an active response of filter feeders that increase mixing thus erasing odor more effectively~\cite{alvarez2018modeling}. Whether and how olfactory shielding may drive the evolution of different locomotion strategies or the emergence of active responses in swimming organisms remains to be tested.  

\section*{Data Availability}
The data that support the findings of this article are openly available~\cite{james_2026_18348686}. The direct numerical simulations are conducted using a modified, GPU-accelerated version of the code available at \url{https://gitlab.com/vdv9265847/IBbookVdV}~\cite{Verzicco2025IBM}.

\section*{Author Contributions}
MJ and AS designed the research. FV wrote the DNS code, which MJ adapted for the present study. MJ conducted the simulations and analyzed the data. All authors contributed to the interpretation of the results. MJ and AS wrote the manuscript with input from FV.

\begin{acknowledgments}
This research was supported by grants to AS from the European Research Council under the European Union’s Horizon 2020 research and innovation programme (grant agreement number 101002724 RIDING) and the National Institutes of Health under award number R01DC018789. The European Commission and the other organizations are not responsible for any use that may be made of the information it contains.
\end{acknowledgments} 

\bibliographystyle{unsrt}
\bibliography{biblio}

\clearpage
\pagebreak

\widetext
\begin{center}
\textbf{\large Supporting Information: Effect of swimming mode on shielding of odor traces in turbulence}
\end{center}
%%%%%%%%%% Merge with supplemental materials %%%%%%%%%%
%%%%%%%%%% Prefix a "S" to all equations, figures, tables and reset the counter %%%%%%%%%%
\setcounter{equation}{0}
\setcounter{figure}{0}
\setcounter{table}{0}
\setcounter{page}{1}
\setcounter{section}{0}
\makeatletter
\renewcommand{\theequation}{S\arabic{equation}}
\renewcommand{\thefigure}{S\arabic{figure}}
\renewcommand{\thetable}{S\arabic{table}}
%%%%%%%%%% Prefix a "S" to all equations, figures, tables and reset the counter %%%%%%%%%%

\section{Modeling Swimmer Dynamics}

As noted in the main text, swimmers are modeled as force dipoles oriented along the streamwise direction, represented as $F(\delta(\bs{x}+\bs{r})-\delta(\bs{x}-\bs{r}))\hat{\bs{x}}$. The magnitude of $F$ is computed according to 

\begin{equation}
    F = C_D\rho Au_s^2/2,
    \label{eq:drag}
\end{equation}
where $C_D$ is the drag coefficient, $\rho$ is the density of the fluid, $A$ is the projected area of the swimmer and $u_s$ is the swimmer speed~\cite{granata1991fluid} relative to the surrounding flow. In the simulations where the swimmers are fixed to a grid, we only account for the mean relative velocity and neglect the $\sim 5$\% turbulent fluctuations. For the simulations that take into account the swimmers' kinematics, we account for the relative fluid flow, including the turbulent fluctuations. Following Ref.~\cite{granata1991fluid}, the values of the drag coefficient are given by
\begin{equation*}
    C_D = \frac{24}{Re_s}(1+0.1315 Re_s^{(0.82-0.05 \log_{10}Re_s)} )
\end{equation*}
for $Re_s = 10$ and
\begin{equation*}
    C_D = \frac{24}{Re_s}(1+0.1935 Re_s^{0.6305})
\end{equation*}
for $Re_s = 25$ and $50$. After an initial simulation, the value of $F$ is further adapted to account for the change in local fluid velocity due to the dynamics of pusher-like and puller-like swimmers.

\section{Benchmarking the passive scalar simulations}

\begin{figure*}
\includegraphics[width=0.9\linewidth]{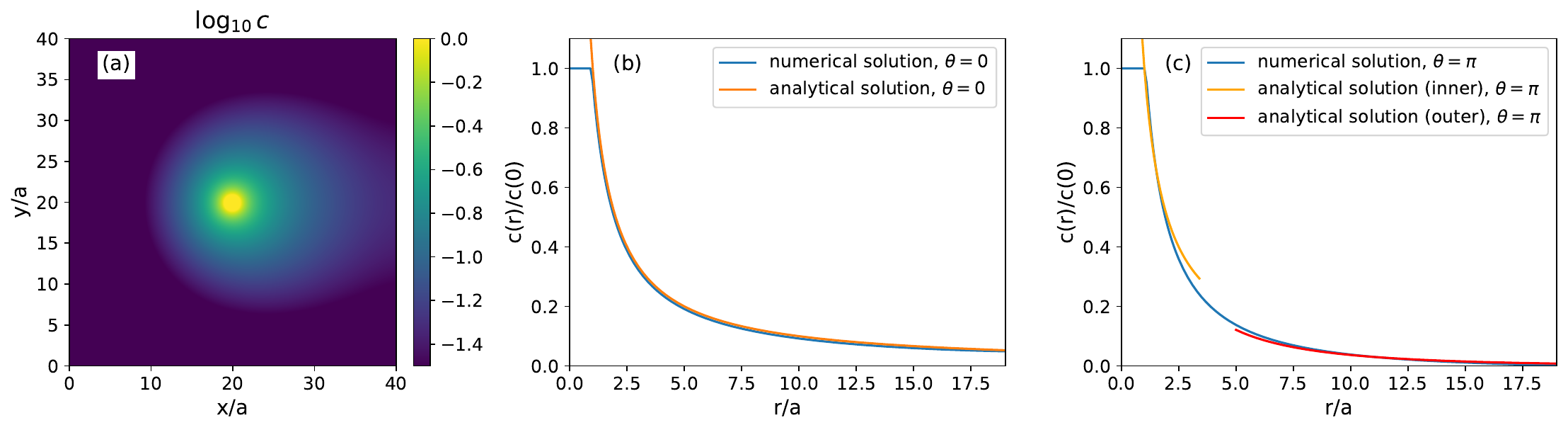} 
\caption{\label{fig:odor_benchmark} Comparison between the numerical solution of the advection-diffusion equation and the corresponding analytical results. (a) Odor field resulting from a sphere of radius $a = 1$, moving through a quiescent fluid. Here $Re = U_\infty a/\nu = 0.1$ and $Pe = U_\infty a/D = 0.1$. (b) Comparison of the simulation results and analytical solution along the wake centerline ($\theta = 0$). (c) Comparison of the simulation results and analytical asymptotics ('inner' and 'outer' solutions) along the front centerline ($\theta = \pi$). }
\end{figure*}

\begin{figure*}
\includegraphics[width=0.9\linewidth]{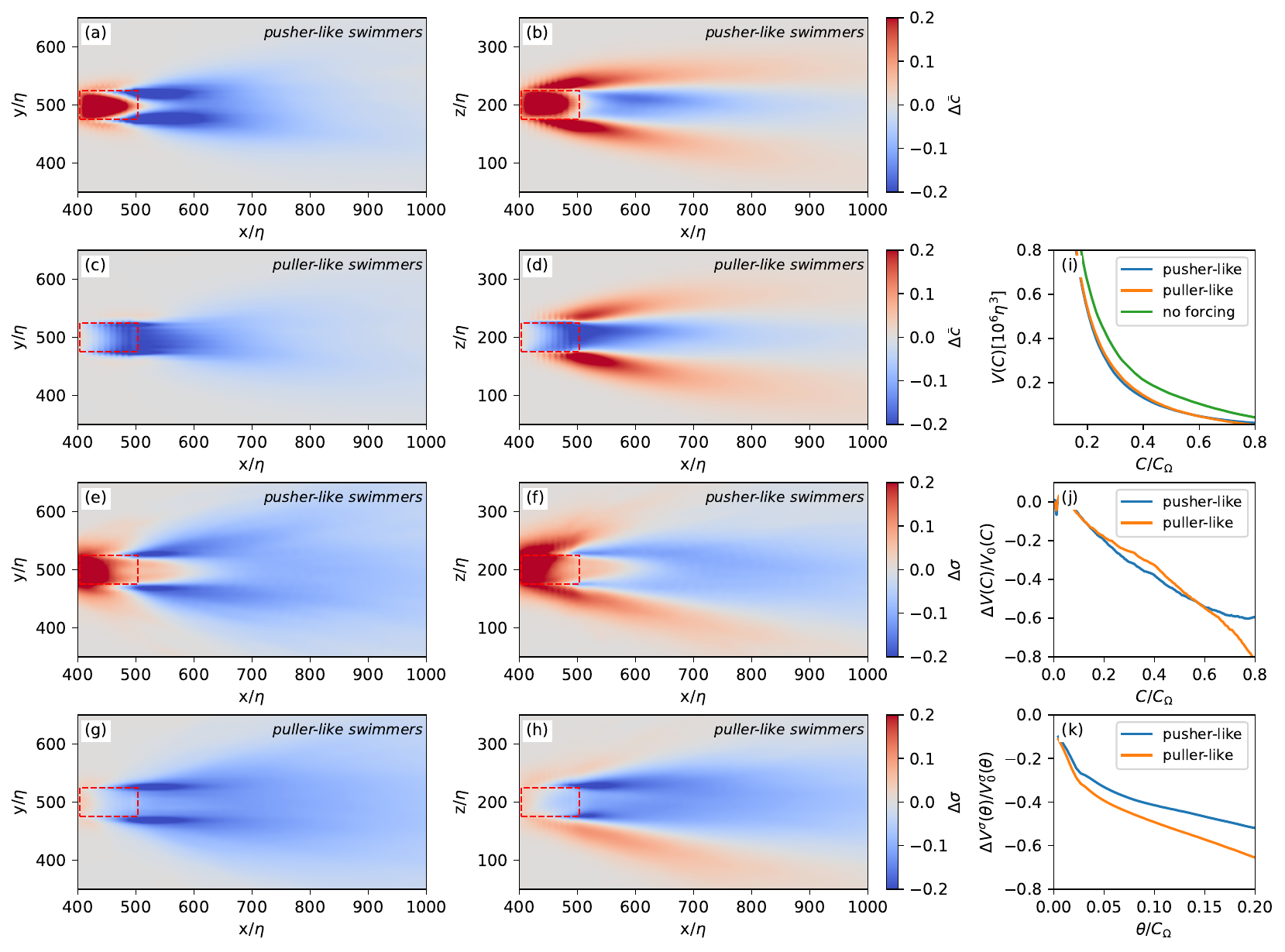} 
\caption{\label{fig:mean_same} Statistics of the mean odor field when the swimmers are arranged in a box of same width to height aspect ratio.  (a) Horizontal and (b) vertical cross-sections of the mean odor field for differential $\Delta \bar{c}(\bs{x})$ for pusher-like swimmers. Panels (c) and (d) show the same for puller-like swimmers. Horizontal and vertical cross-sections of the odor standard deviation differential $\Delta \sigma(\bs x)$ for (e,f) pusher-like and (g,h) puller-like swimmers. (i) Volume $V(C)$ for both types of swimmers and for the baseline case. (j) The corresponding shielding intensities $\Delta V(C)/V_0(C)$. (k) Shielding intensity of the odor fluctuations $\Delta V_\sigma(\theta)/V_{\sigma 0}(\theta)$.} 
\end{figure*}

To benchmark the advection-diffusion equation solver, we compare the simulation results for the odor emitted from a sphere moving through a quiescent fluid with analytical asymptotic solutions. A sphere of radius $a = 1$, implemented using an immersed boundary method~\cite{viola2020fluid}, is placed in a uniform flow. We set Reynolds number to $Re = U_\infty a/\nu = 0.1$ and Peclet number to $Pe = U_\infty a/D = 0.1$, where $U_\infty$ is the flow velocity far from the sphere. The odor field is fixed at $c(0) = c(r=a) = 1$ inside the sphere. At low $Re$ and $Pe$, the leading order terms in the analytical asymptotic expansion for the concentration field near the sphere (inner solution) $c_0(r,\theta)$ and far from the sphere (outer solution) $C_0(r,\theta)$ are given by~\cite{acrivos1962heat}
\begin{align*}
    c_0(r) =& \frac{1}{r} \\
    C_0(r,\theta) =& \frac{1}{r}e^{\epsilon r(\text{cos}(\theta)-1)}.
\end{align*}
Here $\epsilon = Pe$ and $\theta$ is the angle from the streamwise axis. Fig.~\ref{fig:odor_benchmark} shows the comparison between the numerical solution and the analytical asymptotics for $\theta = 0$ and $\pi$ showing excellent agreement. Note that for $\theta = 0$, both $c_0$ and $C_0$ collapses on the same curve.

\begin{figure*}
\includegraphics[width=0.9\linewidth]{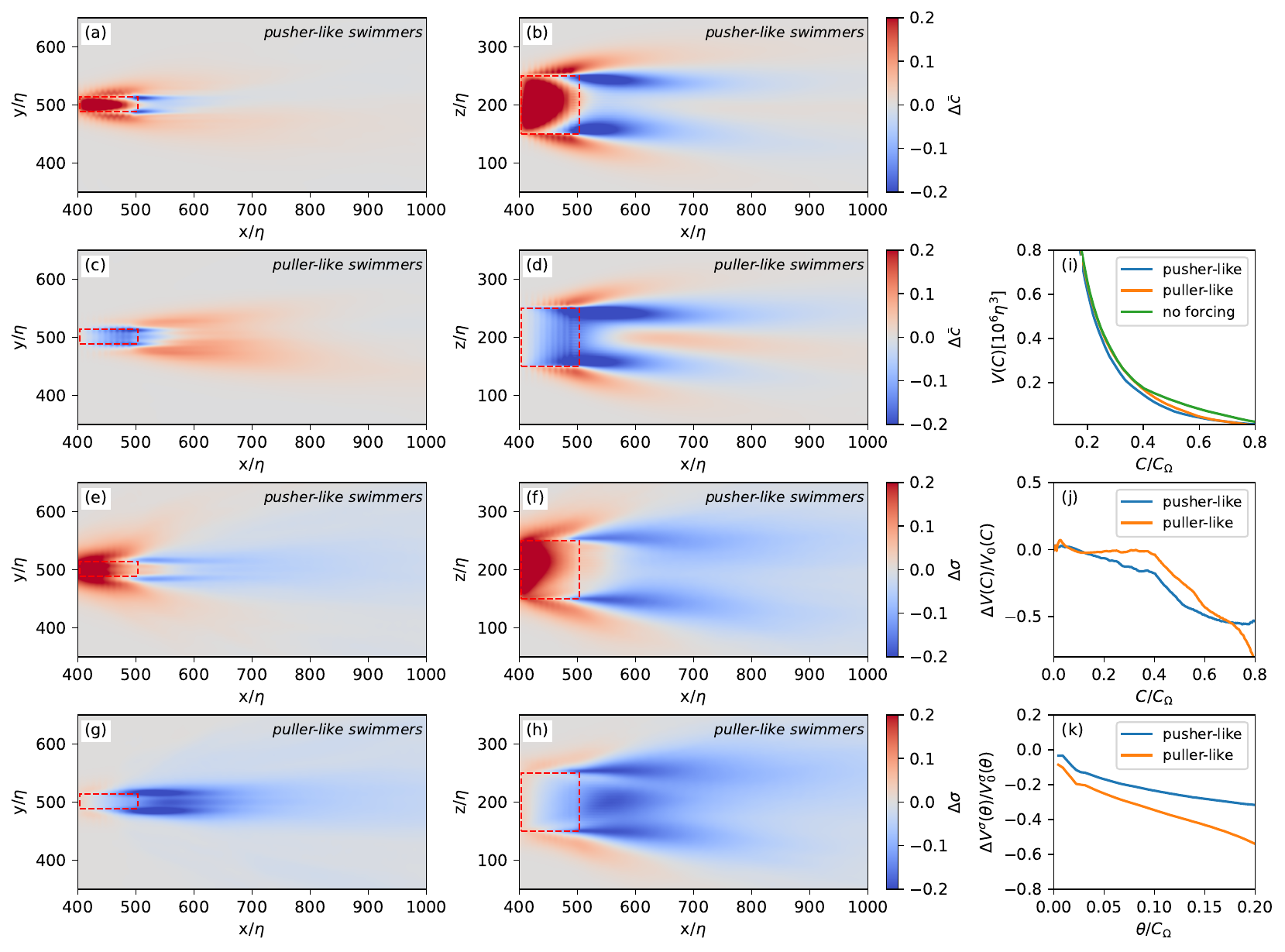} 
\caption{\label{fig:mean_vertical} Statistics of the mean odor field when the swimmers are arranged in a box with a different aspect ratio than those presented in the main text.  Here, the cuboidal box containing the swimmers are vertically oriented, as opposed to the results presented in Fig.~\ref{fig:mean}. Horizontal and (b) vertical cross-sections of the mean odor field for differential $\Delta \bar{c}(\bs{x})$ for pusher-like swimmers. Panels (c) and (d) show the same for puller-like swimmers. Horizontal and vertical cross-sections of the odor standard deviation differential $\Delta \sigma(\bs x)$ for (e,f) pusher-like and (g,h) puller-like swimmers. (i) Volume $V(C)$ for both types of swimmers and for the baseline case. (j) The corresponding shielding intensities $\Delta V(C)/V_0(C)$. (k) Shielding intensity of the odor fluctuations $\Delta V_\sigma(\theta)/V_{\sigma 0}(\theta)$.}
\end{figure*}

\begin{figure*}
\includegraphics[width=0.9\linewidth]{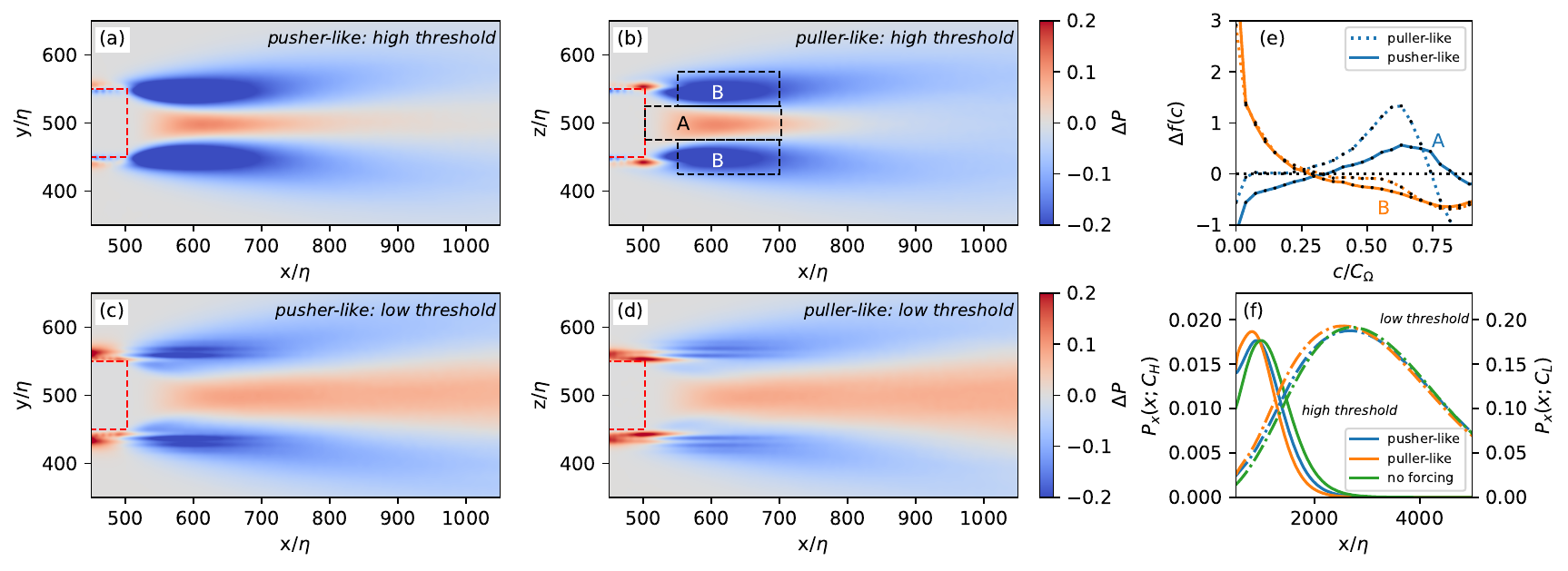} 
\caption{\label{fig:prob_full}  The detection probability differential $\Delta P(\bs x; C_H)$ at a high threshold ($C_H$) for (a) pusher-like and (b) puller-like swimmers. $\Delta P(\bs x; C_L)$ at a low threshold ($C_L$) for (c) pusher-like and (d) puller-like swimmers. (e) Odor pdf differential $\Delta f(c; \Omega_{A/B})$ for pusher-like and puller-like swimmers showing that the hydrodynamic fluctuations due to the swimming dynamics result in an effective eddy diffusivity. (f) Gross detection probability $P_x(x;C)$ as a function of stream-wise direction for both low and high detection thresholds showing that at large distances, the detection probability is damped due to the swimming dynamics.} 
\end{figure*}

\begin{figure*}
\includegraphics[width=1\linewidth]{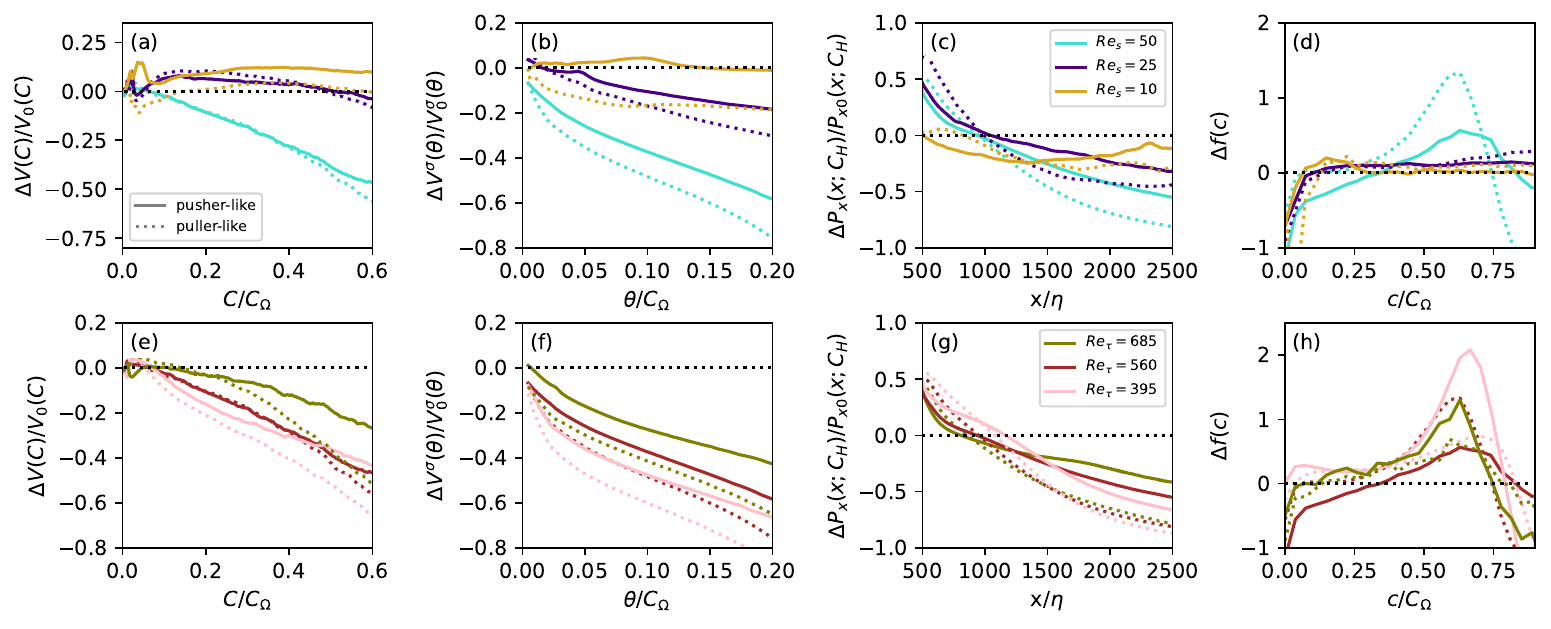} 
\caption{\label{fig:reynolds} (a) Shielding intensity of mean odor $\Delta V(C)/V_0(C)$ as a function of normalized threshold $C/C_\Omega$, (b) shielding of odor fluctuations $\Delta V^\sigma(\theta)/V^\sigma_0(\theta)$ as a function of normalized threshold $s/C_\Omega$, (c) normalized detection probability differential $\Delta P_x(x;C_H)/P_x0(x;C_H)$ as a function of the streamwise distance from the swimmers and (d) odor pdf differential $\Delta f(c; \Omega_A/B)$. Colors represent different swimmer Reynolds numbers (a-d) or different turbulent Reynolds numbers (e-h). Pusher-like and puller-like swimmers are represented by solid and dotted lines respectively.}
\end{figure*}

\begin{figure*}
\includegraphics[width=0.9\linewidth]{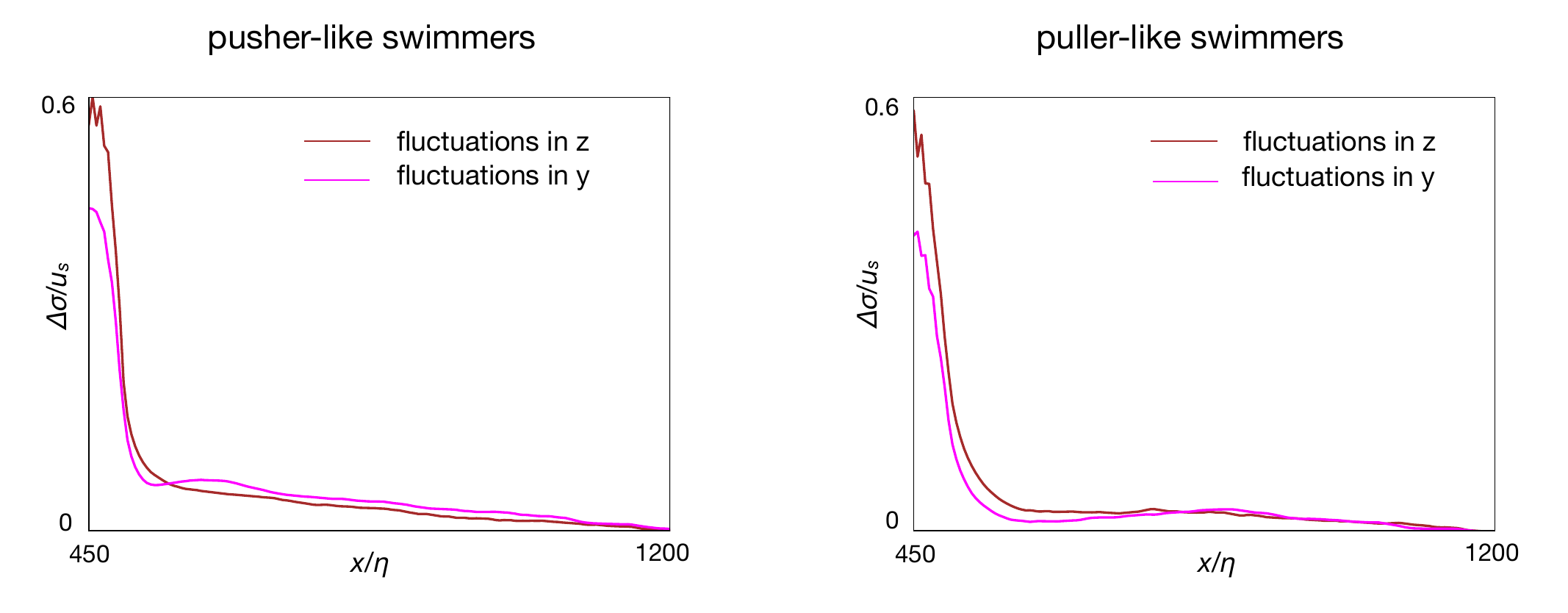} 
\caption{\label{fig:vel_si}  Differences in the standard deviation of the velocity fluctuations in the wall-normal and span-wise directions for pusher-like and puller-like swimmers.} 
\end{figure*}

\begin{figure*}
\includegraphics[width=0.6\linewidth]{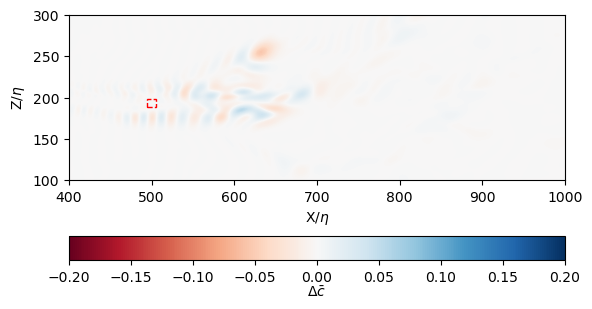} 
\caption{\label{fig:single_si}  Mean odor shielding by an individual pusher-like swimmer located in the red region (Environment 1), showing that a single swimmer does not yield a significant shielding effect.} 
\end{figure*}

\begin{figure*}
\includegraphics[width=0.75\linewidth]{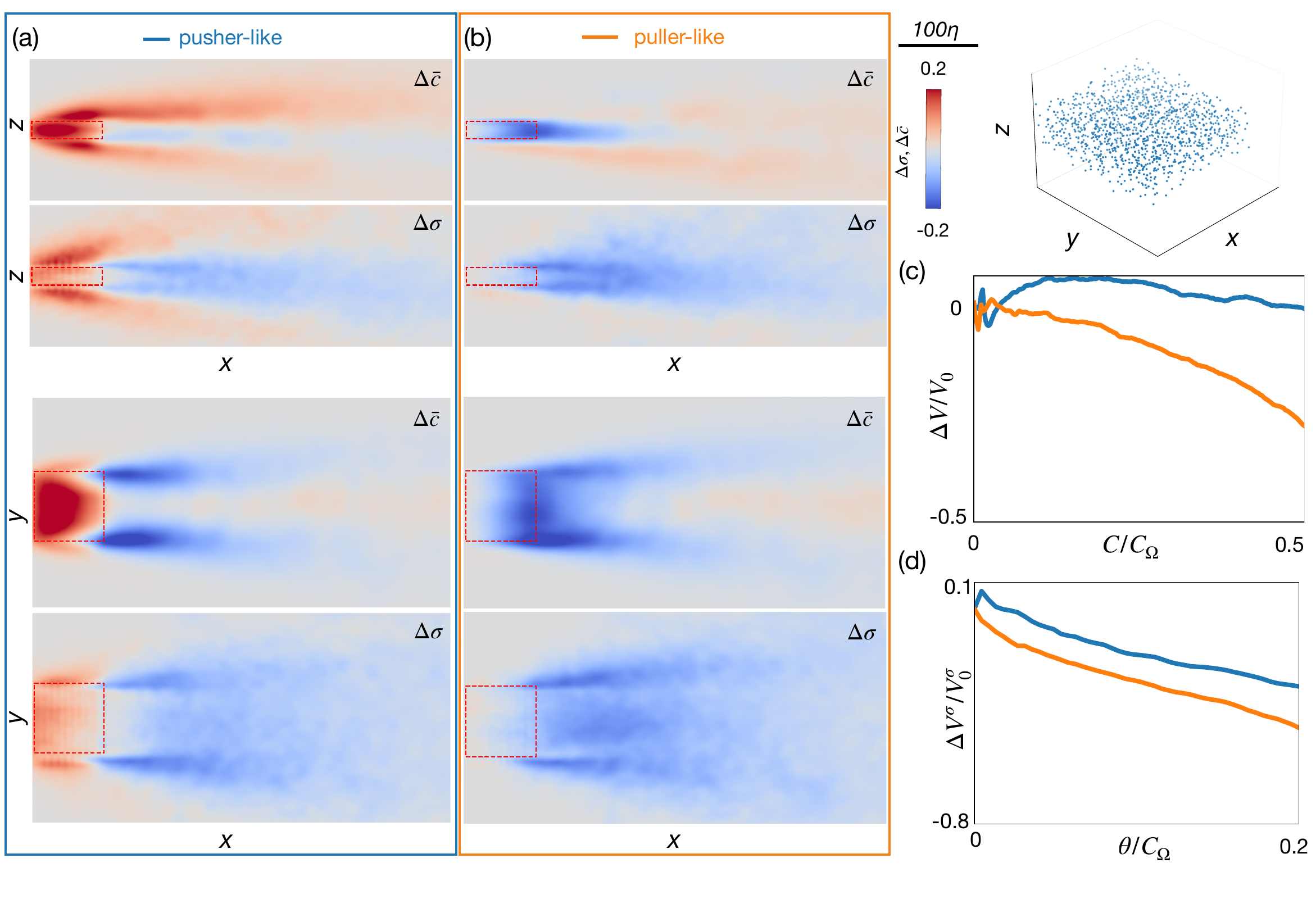} 
\caption{\label{fig:kinematics_low_noise} Effect of swimmers' kinematics on odor shielding for an intermediate value of positional noise ($\sigma_{MS}\approx d$). A snapshot of the swimmers' positions is shown on the top right. Vertical and horizontal cross-sections of $\Delta \bar{c}(\bs{x})$ and $\Delta \sigma(\bs x)$ for (a) pusher-like swimmers and (b) puller-like swimmers. (c) Shielding intensities of the mean odor field $\Delta V(C)/V_0(C)$. (d) Shielding intensity of the odor fluctuations $\Delta V^\sigma(\theta)/V^\sigma_0(\theta)$ for pusher-like and puller-like swimmers.}
\end{figure*}

\begin{figure*}
\includegraphics[width=0.75\linewidth]{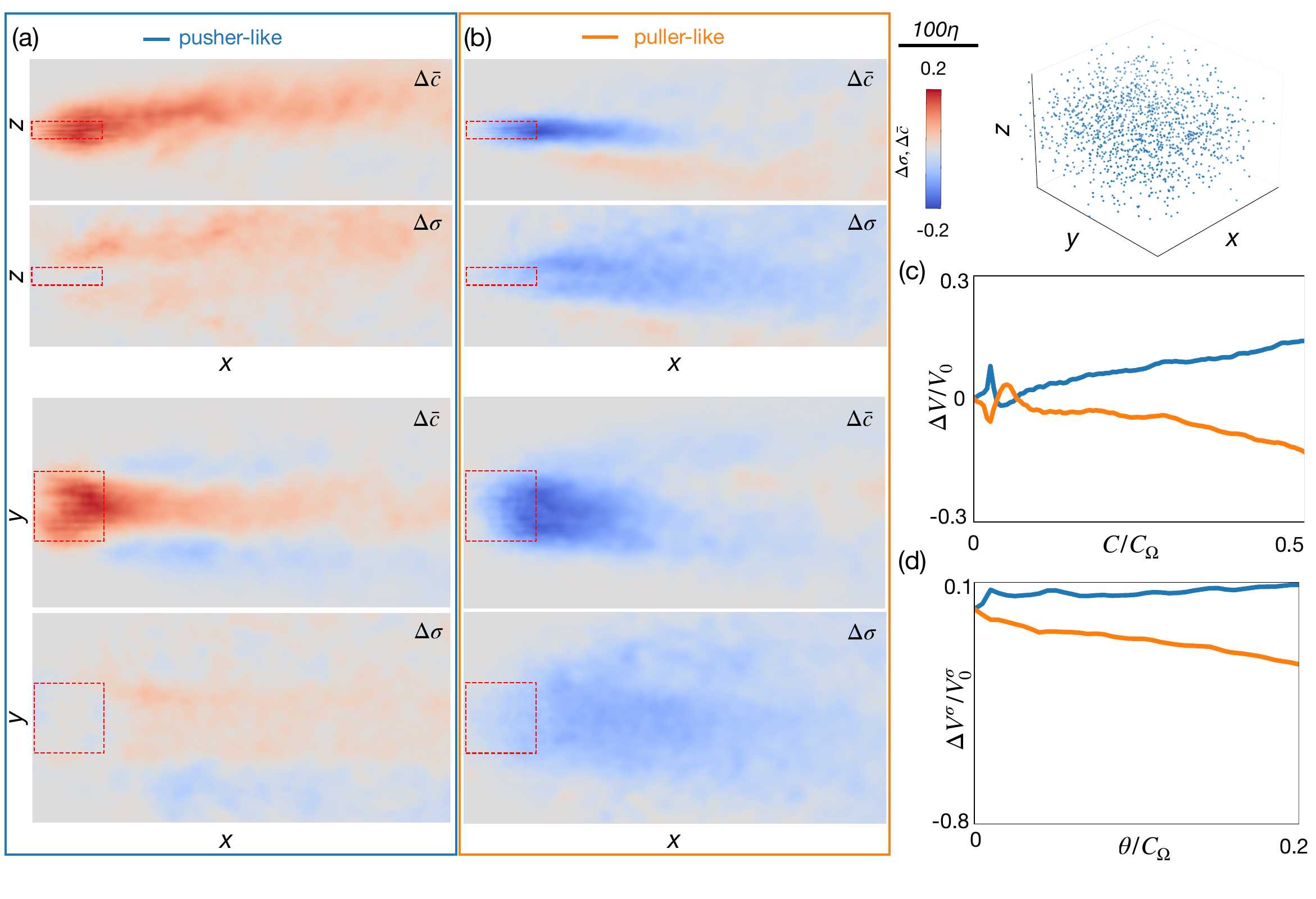} 
\caption{\label{fig:kinematics_high_noise} Effect of swimmers' kinematics on odor shielding for a high value of positional noise ($\sigma_{MS}\approx 5 d$). A snapshot of the swimmers' positions is shown on the top right. Vertical and horizontal cross-sections of $\Delta \bar{c}(\bs{x})$ and $\Delta \sigma(\bs x)$ for (a) pusher-like swimmers and (b) puller-like swimmers. (c) Shielding intensities of the mean odor field $\Delta V(C)/V_0(C)$. (d) Shielding intensity of the odor fluctuations $\Delta V^\sigma(\theta)/V^\sigma_0(\theta)$ for pusher-like and puller-like swimmers. }
\end{figure*}

\end{document}